\documentstyle[twoside,epsf]{article}

\catcode`\@=11
\long\def\@makefntext#1{
\protect\noindent \hbox to 3.2pt {\hskip-.9pt  
$^{{\eightrm\@thefnmark}}$\hfil}#1\hfill}		

\def\@makefnmark{\hbox to 0pt{$^{\@thefnmark}$\hss}}	
	
\def\ps@myheadings{\let\@mkboth\@gobbletwo
\def\@oddhead{\hbox{}
\rightmark\hfil\eightrm\thepage}   
\def\@oddfoot{}\def\@evenhead{\eightrm\thepage\hfil
\leftmark\hbox{}}\def\@evenfoot{}
\def\sectionmark##1{}\def\subsectionmark##1{}}

\oddsidemargin=\evensidemargin
\newcounter{sectionc}\newcounter{subsectionc}\newcounter{subsubsectionc}
\renewcommand{\section}[1] {\vspace{12pt}\addtocounter{sectionc}{1} 
\setcounter{subsectionc}{0}\setcounter{subsubsectionc}{0}\noindent 
	{\tenbf\thesectionc. #1}\par\vspace{5pt}}
\renewcommand{\subsection}[1] {\vspace{12pt}\addtocounter{subsectionc}{1} 
\setcounter{subsubsectionc}{0}\noindent 
{\bf\thesectionc.\thesubsectionc. {\kern1pt \bfit #1}}\par\vspace{5pt}}
\renewcommand{\subsubsection}[1] {\vspace{12pt}\addtocounter{subsubsectionc}{1}
	\noindent{\tenrm\thesectionc.\thesubsectionc.\thesubsubsectionc.
	{\kern1pt \tenit #1}}\par\vspace{5pt}}
\newcommand{\nonumsection}[1] {\vspace{12pt}\noindent{\tenbf #1}
	\par\vspace{5pt}}
	
\newcommand{\eqref}[1] {(\ref{#1})}		

\newcounter{appendixc}
\newcounter{subappendixc}[appendixc]
\newcounter{subsubappendixc}[subappendixc]
\renewcommand{\thesubappendixc}{\Alph{appendixc}.\arabic{subappendixc}}
\renewcommand{\thesubsubappendixc}
	{\Alph{appendixc}.\arabic{subappendixc}.\arabic{subsubappendixc}}

\renewcommand{\appendix}[1] {\vspace{12pt}
        \refstepcounter{appendixc}
        \setcounter{figure}{0}
        \setcounter{table}{0}
        \setcounter{lemma}{0}
        \setcounter{theorem}{0}
        \setcounter{corollary}{0}
        \setcounter{definition}{0}
        \setcounter{equation}{0}
        \renewcommand{\thefigure}{\Alph{appendixc}.\arabic{figure}}
        \renewcommand{\thetable}{\Alph{appendixc}.\arabic{table}}
        \renewcommand{\theappendixc}{\Alph{appendixc}}
        \renewcommand{\thelemma}{\Alph{appendixc}.\arabic{lemma}}
        \renewcommand{\thetheorem}{\Alph{appendixc}.\arabic{theorem}}
        \renewcommand{\thedefinition}{\Alph{appendixc}.\arabic{definition}}
        \renewcommand{\thecorollary}{\Alph{appendixc}.\arabic{corollary}}
        \renewcommand{\theequation}{\Alph{appendixc}.\arabic{equation}}
        \noindent{\tenbf Appendix \theappendixc #1}\par\vspace{5pt}}
\newcommand{\subappendix}[1] {\vspace{12pt}
        \refstepcounter{subappendixc}
        \noindent{\bf Appendix \thesubappendixc. {\kern1pt \bfit #1}}
	\par\vspace{5pt}}
\newcommand{\subsubappendix}[1] {\vspace{12pt}
        \refstepcounter{subsubappendixc}
        \noindent{\rm Appendix \thesubsubappendixc. {\kern1pt \tenit #1}}
	\par\vspace{5pt}}

	\newcommand{\ket}[1]{\left | #1 \right\rangle}
	\newcommand{\bra}[1]{\left \langle #1 \right |}
	
	\newcommand{\tr}{\mbox{tr}}
	\newcommand{\M}{{\mathcal{M}}}
	\newcommand{\K}{{\mathcal{K}}}
	\newcommand{\C}{{\mathcal{C}}}
	\newcommand{\Q}{{\mathcal{Q}}}
	\newcommand{\T}{{\mathcal{T}}}
	\newcommand{\LL}{{\mathcal{L}}}
	\newcommand{\ep}{\epsilon}

\newcommand{\textlineskip}{\baselineskip=13pt}
\newcommand{\smalllineskip}{\baselineskip=10pt}

\newcommand{\copyrightheading}[1]
	{\vspace*{-2.5cm}\smalllineskip{\flushleft
	{\footnotesize Quantum Information and Computation, Vol.~1, No.~0 (2001) 000--000 #1}\\
	{\footnotesize \copyright\kern2pt Rinton Press}\\
	 }}

\newcommand{\publisher}[2]{{\begin{center}\footnotesize\smalllineskip 
	Received #1\\
	Revised #2
	\end{center}
	}}

\def\abstracts#1#2#3{{
	\centering{\begin{minipage}{4.5in}\footnotesize\baselineskip=10pt
	\parindent=0pt #1\par 
	\parindent=15pt #2\par
	\parindent=15pt #3
	\end{minipage}}\par}} 

\def\keywords#1{{
	\centering{\begin{minipage}{4.5in}\footnotesize\baselineskip=10pt
	{\footnotesize\it Keywords}\/: #1
	 \end{minipage}}\par}}
\def\communicate#1{{
	\centering{\begin{minipage}{4.5in}\footnotesize\baselineskip=10pt
	{\footnotesize\it Communicated by}\/: #1
	 \end{minipage}}\par}}

\newcounter{itemlistc}
\newcounter{romanlistc}
\newcounter{alphlistc}
\newcounter{arabiclistc}

\newcommand{\fcaption}[1]{
        \refstepcounter{figure}
        \setbox\@tempboxa = \hbox{\footnotesize Fig.~\thefigure. #1}
        \ifdim \wd\@tempboxa > 5in
           {\begin{center}
        \parbox{5in}{\footnotesize\smalllineskip Fig.~\thefigure. #1}
            \end{center}}
        \else
             {\begin{center}
             {\footnotesize Fig.~\thefigure. #1}
              \end{center}}
        \fi}

\newcommand{\tcaption}[1]{
        \refstepcounter{table}
        \setbox\@tempboxa = \hbox{\footnotesize Table~\thetable. #1}
        \ifdim \wd\@tempboxa > 5in
           {\begin{center}
        \parbox{5in}{\footnotesize\smalllineskip Table~\thetable. #1}
            \end{center}}
        \else
             {\begin{center}
             {\footnotesize Table~\thetable. #1}
              \end{center}}
        \fi}

\def\pmb#1{\setbox0=\hbox{#1}
	\kern-.025em\copy0\kern-\wd0
	\kern.05em\copy0\kern-\wd0
	\kern-.025em\raise.0433em\box0}

\def\fnt#1#2{\footnotetext{\kern-.3em
	{$^{\mbox{\scriptsize #1}}$}{#2}}}

\def\fpage#1{\begingroup
\voffset=.3in
\thispagestyle{empty}\begin{table}[b]\centerline{\footnotesize #1}
	\end{table}\endgroup}

\def\runninghead#1#2{\pagestyle{myheadings}
\markboth{{\protect\footnotesize\it{\quad #1}}\hfill}
{\hfill{\protect\footnotesize\it{#2\quad}}}}
\font\tenrm=cmr10
\font\tenit=cmti10 
\font\tenbf=cmbx10
\font\bfit=cmbxti10 at 10pt
\font\ninerm=cmr9

\font\eightrm=cmr8

\newtheorem{theorem}{\noindent Theorem}

\newtheorem{proposition}{\noindent Proposition}

\newtheorem{lemma}{\noindent Lemma}

\newtheorem{definition}{\noindent Definition}
\newtheorem{corollary}{\indent Corollary}

\newcommand{\proof}[1]{{\noindent \bf Proof.} #1 $\Box$.}

\def\FigName{figure}%
\newbox\captionbox
\long\def\@makecaption#1#2{%
  \ifx\FigName\@captype
    \vskip\abovecaptionskip
    \setbox\tempbox\hbox{{\figurecaptionfont #1\hskip1em #2}}
	\ifdim\wd\tempbox< 28pc
	\centerline{\box\tempbox}
	\else.\pi_\rho U
	{\figurecaptionfont #1\hskip1em #2\par}
\fi\else
  	\setbox\tempbox\hbox{{\tablecaptionfont #1\hskip1em #2}}
 	\ifdim\wd\tempbox< 28pc 
	\centerline{\box\tempbox}
	\else
	{\tablecaptionfont #1\hskip1em #2\par}%
	\fi   
 \vskip\belowcaptionskip
 \fi}
\InputIfFileExists{psfig.sty}
{\typeout{^^Jpsfig.sty inputed...ok}}{\typeout{^^JWarning: psfig.sty could be be found.^^J}}
\InputIfFileExists{epsfsafe.tex}
{\typeout{^^Jepsfsafe.tex inputed...ok}}
			{\typeout{^^JWarning: epsfsafe.tex could not be found.^^J}}
\InputIfFileExists{epsfig.sty}
{\typeout{^^Jepsfig.sty inputed...ok}}{\typeout{^^JWarning: epsfig.sty could not be found.^^J}}
\InputIfFileExists{epsf.sty}
{\typeout{^^Jepsf.sty inputed...ok}}{\typeout{^^JWarning: epsf.sty could not be found.^^J}}%
\def\fps@figure{tbp}
\def\ftype@figure{1}
\def\ext@figure{lof}
\def\fnum@figure{Fig.\ \thefigure}
\def\qed{\hbox{${\vcenter{\vbox{	          
   \hrule height 0.4pt\hbox{\vrule width 0.4pt height 6pt
   \kern5pt\vrule width 0.4pt}\hrule height 0.4pt}}}$}}

\begin{document}
\setlength{\textheight}{8.0truein}    

\runninghead{Criteria for measures of quantum correlations} 
            {A. Brodutch and K. Modi}

\normalsize\textlineskip
\thispagestyle{empty}
\setcounter{page}{1}

\copyrightheading{}	

\vspace*{0.88truein}

\fpage{1}
\centerline{\bf
Criteria for measures of quantum correlations}
\vspace*{0.37truein}
\centerline{\footnotesize 
Aharon Brodutch\footnote{cap.fwiffo@gmail.com}}
\vspace*{0.015truein}
\centerline{\footnotesize\it Department of Physics \& Astronomy, Faculty of Science, Macquarie University}
\baselineskip=10pt
\centerline{\footnotesize\it Sydney, NSW 2109, Australia}
\vspace*{10pt}
\centerline{\footnotesize Kavan Modi\footnote{kavan@quantumlah.org}}
\vspace*{0.015truein}
\centerline{\footnotesize\it Department of Physics, University of Oxford, Clarendon Laboratory, Oxford, UK}

\centerline{\footnotesize\it and}

\centerline{\footnotesize\it Centre for Quantum Technologies, National University of Singapore}
\baselineskip=10pt
\centerline{\footnotesize\it Singapore}
\vspace*{0.225truein}
\publisher{(received date)}{(revised date)}

\vspace*{0.21truein}
\abstracts{
Entanglement does not describe all quantum correlations and several authors have shown the need to go beyond entanglement when dealing with mixed states. Various  different measures have sprung up in the literature, for a variety of reasons, to describe bipartite and multipartite quantum correlations; some are known under the collective name {\it quantum discord}.  Yet, in the same sprit as the criteria for entanglement measures, there is no general mechanism that determines whether a measure of quantum and classical correlations is a proper measure of correlations. This is partially due to the fact that the answer is a bit muddy. In this article we attempt tackle this muddy topic by writing down several criteria for a ``good" measure of correlations. We breakup our list into \emph{necessary}, \emph{reasonable}, and \emph{debatable} conditions. We then proceed to prove several of these conditions for generalized measures of quantum correlations. However, not all conditions are met by all measures; we show this via several examples. The reasonable conditions are related to continuity of correlations, which has not been previously discussed. Continuity is an important quality if one wants to probe quantum correlations in the laboratory. We show that most types of quantum discord are continuous but none are continuous with respect to the measurement basis used for optimization. }{}{}

\vspace*{10pt}
\vspace*{3pt}
\communicate{to be filled by the Editorial}

\vspace*{1pt}\textlineskip	
\section{Introduction}       
\vspace*{-0.5pt}
\noindent
Quantum systems can have correlations different than those of classical systems. These correlations act as a resource for some tasks and an obstacle for others. Examples of the tasks that benefit from  quantum correlations are quantum teleportation of unknown states~\cite{PhysRevLett.70.1895} and dense coding~\cite{densecoding}. The speed up associated with quantum computation has often been attributed to the ability to create stronger than classical correlations~\cite{speedup}. On the other hand, quantum correlations are also responsible for decoherence and dissipation of quantum systems~\cite{breuer02a}. Further, the quantum nature of multipartite systems makes some tasks very hard or even impossible; the examples range from the impossibility to distinguish ensembles using local operations and classical communications (LOCC)~\cite{NLWE} to the impossibility of implementing most quantum gates using LOCC~\cite{BTgates}. Some of these obstacles can be overcome by using correlated quantum systems, while others remain impossible. For example the instantaneous non-demolition measurements of some nonlocal observables violates causality~\cite{PhysRevD.34.1805}. Most of these obstacles can be used as an advantage for some tasks such as cryptography.

The difference between correlations in quantum and classical systems tempt us to divide correlations into quantum and classical parts. Recently, a class of measures for quantum, classical, and total correlations based on the effects of measurements on a system have been explored~\cite{arXiv:quant-ph/0011039, hv, arXiv:quant-ph/0105072, arXiv:quant-ph/0112074, PhysRevA.66.022104, PhysRevA.72.032317, arXiv:0707.2195, PhysRevA.77.022301, PhysRevA.83.012312, arXiv:0905.2123, arXiv:0911.5417, arXiv:1004.0190, arXiv:1005.4348, MINL}, often these measures are collective named \emph{quantum discord}. A great deal of work has been done on quantum discord and related measures of correlations in the last decade (see~\cite{arXiv:1112.6238,arXiv:1107.3428} and references within). Unlike the earlier work on this subject, these measures account for quantum correlations other than entanglement. They can also be extended easily to quantify multipartite correlations. Many of these measures  have been related to various tasks in quantum information and quantum computation, e.g. The DQC1 model ~\cite{KL, DattaShajiCaves, PhysRevA.84.012313}, no broadcasting~\cite{arXiv:0707.0848, arXiv:0901.1280, LMP-Luo}, quantum metrology~\cite{arXiv:1003.1174}, quantum state merging~\cite{merging, PhysRevA.83.032324, arXiv:1008.4135}, and quantum thermodynamics~\cite{arXiv:quant-ph/0202123, BTdemons}. There have also been several proposal to experimentally detect discord and other experimental investigations relating to discord~\cite{arXiv:0807.0668, arXiv:0911.2848, arXiv:1001.5441, arXiv:0911.3460, arXiv:1102.4710, arXiv:1104.1596, arXiv:1104.3885}.

While some of these measures have an operational interpretation in terms of  various tasks~\cite{arXiv:0707.0848, arXiv:0901.1280, PhysRevA.83.032324, arXiv:1008.4135, PhysRevLett.106.160401, PhysRevLett.106.220403, arXiv:1105.2768}, most of the work on the subject did not include an in-depth analysis of the correlation measures from an information theoretic prospective. Most measures are not judged according to any information theoretic criteria like the criteria for entanglement measures~\cite{arXiv:quant-ph/0504163}. The exception is the seminal paper by Henderson and Vedral~\cite{hv} where five different criteria  for a measure of classical information were presented. The Henderson-Vedral measure of classical correlations was designed to meet four of those criteria, namely: (a) product states are uncorrelated, (b) classical correlations are invariant under local unitary operations, (c) classical correlations are non increasing under local operations (without communication), and (d) for pure states the classical correlations are the entropy of the reduced local states. The fifth criteria, symmetry under the interchange of the local subsystems was conjunctured but was later found to be inconsistent with their measure.

In this article we present a general method for creating meaningful discord-like measures of classical, quantum, and total correlations. In the spirit of~\cite{PhysRevA.72.032317} the quantum and classical correlations are defined via some change in the state before and after a measurement.  We present a set of criteria which could be applied to test these measures and discuss the various known measures in terms of these criteria. We divide these criteria into different categories: Necessary; Reasonable and Debatable, and suggest that any measure which does not meet the necessary criteria  is not a valid measure of correlations. The importance of the remaining criteria is discussed but we leave the questions of which ones should be adopted and which should be dropped open. The reasonable conditions are related to continuity, a property which has so far not been discussed in relation to quantum discord. We explore this property in detail, proving continuity for some measures and showing discontinuity for others, we also discuss the role of continuity in the scenario of Maxwell's demons extracting work from a heat bath.

The paper is organized as follows: In Sec. 2  we define a generalized function, set of measurements, and measurement strategies that give a measure of generalized correlations. In Sec. 3 we lay out further constrains on generalized correlations measures. In the following three Sections we discuss these conditions in detail. In Sec. 7 we discuss some standard measures and whether they follow the conditions of Sec. 3. In Sec. 8 we discuss the role of continuity using  the important case of Maxwell's demon in light of our results followed by the conclusions in Sec. 9.

\section{Quantum, classical and total correlations}
\noindent
Quantum discord is often used as a measure of quantum correlations. There are several other schemes  for  quantifying quantum correlations  (see Sec.~7), e.g. \emph{measurement induced disturbance} (MID) and geometric discord (based on the Hilbert-Schmidt distance). Each of these schemes require different methods of identifying the total, quantum, and classical correlations. Noting the commonalities among the different schemes we can quantify correlations in the most general sense via a positive-real valued function and a set local measurements.

{\definition The generalized-discord function $\K[\rho_{1},\rho_{2}]$ is defined over all quantum states $\rho_1$ and $\rho_2$ with properties $\K[\rho_1,\rho_2]\in \mathsf{R}^+$ and $\K[\rho,\rho]=0$.}

{\definition The set of measurements $\{\M\}$ is a set of general quantum-operations that are trace preserving (complete).}

{\definition $\M(\rho)$ is a classically correlated state.}

Classical states can be defined through $\M(\rho)$ or the other way around. Usually $\{\M\}$ is a full set of \emph{positive operator values measurements} (POVM) on one or more of the subsystems \emph{with or without} communication. However, most of the time we will only need only local rank-1 POVM or orthogonal measurements, which are included. This is not essential for most of our general claims as long as the properties of Def. 2 are satisfied.

The measurement will typically depend on the quantum state $\rho$, and  is therefore labeled as $\M_\rho$. The method for choosing $\M_\rho$ can be classified into four different strategies.
\begin{itemize}
 \item S1: The measurement is the same for all $\rho$ so $\{\M\}=\M_\chi$.
 \item S2q: $\M_\rho$ will minimize the quantum correlations.
 \item S2c: $\M_\rho$ will maximize classical correlations. 
 \item S3: The measurement does not change the marginals $\M_\rho(\pi_\rho)=\pi_\rho$. 
\end{itemize}
The subscript $\chi$ represents a fixed basis for S1 and a subscript $\rho$ says that $\M$ depends on the state.

Within this scheme there are four fundamental states that we have to deal with~\cite{modivedral}: $\rho$ ,  $\M(\rho)$, and their marginals: $\pi_\rho\equiv\bigotimes_i\rho^i$ and $\M(\pi_\rho) \equiv \bigotimes_i[\M(\rho)]^i$, where $\rho^i=\tr_{\bar{i}}[\rho]$ are the local marginals. We can now use the generalized-discord function $\K$ to define classical quantum and total correlations. 

\begin{itemize}
\item The generalized quantum discord (quantum correlations) of a state $\rho$ is defined as the distance\footnote{The term distance is to be used lightly as $\K$ may not be a proper distance.} between the state $\rho$ and the classical state $\M(\rho)$.
\begin{equation}
  \Q(\rho)\equiv\K[\rho,\M_\rho(\rho)].
\end{equation}
\item Classical correlations of a state $\rho$ are defined as the distance between the classical state $\M(\rho)$ and the reduced product state after the same measurement $\M(\pi_\rho)$.
\begin{equation}
  \C(\rho)\equiv\K[\M(\rho),\M(\pi_\rho)].
\end{equation}
\item The total correlations of a state $\rho$ are defined as the distance between that state $\rho$ and the reduced product $\pi_\rho$
 \begin{equation}
  \T(\rho)\equiv\K[\rho,\pi_\rho].
\end{equation}
\item For completeness we can also define a forth quantity 
 \begin{equation}
  \LL(\rho)\equiv\K[\pi_\rho,\M({\pi_\rho})].
\end{equation}
The last equation quantifies the amount of coherence lost in the local states, i.e. $\rho^i$, when a quantum state is measured. This quantity  is always zero for S3 since the measurements do not disturb the local states.
\end{itemize}

\section{General properties for correlations}
\noindent
Whatever method we use correlations must satisfy more than just the basic properties stated above. The purpose of this paper is to comment on these properties. In the spirit of the conditions on entanglement measures~\cite{vp} we present a set of conditions for  correlations independent of any specific function $\K$. These could be re-defined as conditions on $\K$, $\M$ or the measurement strategy. However apart from 2(b) and 2(c) they are general conditions that should be satisfied by any method used to define correlations, including entanglement monotones.  The conditions we present below are divided into three types. The \emph{necessary conditions} are those that cannot be violated by any measure of correlations. The \emph{reasonable conditions}, related to continuity, would be welcome byproducts of any measure even if they are not enforced, especially in light of experimental limitations. The last set of are the \emph{debatable conditions}; they arrive from some preconceived notions about the nature of correlations. Most of the conditions presented below are based on those of entanglement measures~\cite{vp} and those on classical correlations  defined in~\cite{hv}.
 
\begin{enumerate}
 \item{Necessary conditions}
\begin{enumerate}
 \item Product states have no correlations: $\T(\pi)=\Q(\pi)=\C(\pi)=0$.
 \item All correlations are invariant under local unitary operations. 
 \item All correlations are non-negative: $\C\geq0$ and $\Q\geq0$ and $\T\geq0$.
 \item $\T$ is non-increasing under local operations. 
 \item Classical states have no quantum correlations. $\Q(\M_\chi(\rho))=0$ for all $\rho$ and $\M_\chi$.
\end{enumerate}
 \item{Reasonable conditions}
\begin{enumerate}
 \item Continuity under small perturbations. 
 \item Strong continuity of $\M_\rho$ under small perturbations (SCM).
 \item Weak continuity of $\M_\rho$ under small perturbations (WCM). 
\end{enumerate}
 \item{Debatable conditions}
\begin{enumerate}
 \item For pure bipartite states total, quantum, and classical correlations can be defined by the marginals: $\T(\rho)=G_t(\rho^A)$, $\Q(\rho)=G_q(\rho^A)$, $\C(\rho)=G_c(\rho^A)$.
 \item Correlations are additive: $\T=\C+\Q$.
 \item $\C$ and/or $\Q$ are non increasing under local operations.
 \item Symmetry under interchange of the subsystems.
\end{enumerate}
\end{enumerate}

A detailed discussion of these conditions using  the scheme presented in Sec. 2  follows.

\section{Necessary conditions}\label{NC}

\subsection{No correlation in product states}
\noindent
{\proposition Condition 1(a) is satisfied for $\T$ and $\C$ for all strategies.}

\proof{By definition of $\K$, $\T(\pi)=\K[\pi,\pi]=0$ and $\C(\pi)=\K[\M(\pi),\M(\pi)]=0$. However, if $\T$ is defined differently we must be careful as these condition may be violated}

{\proposition For strategies S2q, S2c, and S3 Condition 1(a) for $\Q$ is satisfied.}

\proof{Let us begin with S3, where $\M$ is a projective measurement in the basis of $\pi$ with $\M(\pi)=\pi$. The value of quantum correlations is $\Q(\pi)=\K[\pi,\pi]=0$, which minimizes $\Q$ thus satisfying the criteria S2q. For S2c $\C$ is always 0 due to Prop. 1 for any $\M$. Thus, if $\M$ must be chosen carefully, we may choose the one satisfying $\M(\pi)=\pi$, then $\Q$ too will vanish}

{\proposition For strategies S1 Condition 1(a) for $\Q$ may fail.}

\proof{We show this by an explicit example. Let us use the Hilbert-Schmidt distance as the function $\K$: $\K[\rho_1,\rho_2]=\tr[(\rho_1-\rho_2)^2]$. Let $\pi$ be diagonal in the computational basis and make $\mathcal{M}$ a measurement in a basis different from the computational basis for each system. Then we get: $\M(\pi)\neq\pi$ and $\Q(\pi) =\tr[(\pi-\M(\pi))^2]>0$}

We note that for discord based on mutual information or conditional entropy condition 1(a) is satisfied since both quantities vanish for product states.

\subsection{Invariance under local unitary}
\noindent
To have invariance under local unitary, $U_L\equiv\bigotimes_i u^i$, we have to restrict ourselves to functions that are invariant under local unitary.

{\definition Function $\K$ is called a unitary-invariant function if it satisfies
\begin{equation}
\K[\rho_1,\rho_2]=\K[u\rho_1 u^\dag, u\rho_2 u^\dag].\label{unitaryK}
\end{equation}}

{\proposition Under Eq.~\eqref{unitaryK} Condition 1(b) for $\T$ is satisfied for all strategies.}

\proof {By Def. 4 of $\K$, $\T(U_L\rho U_L^\dag)=\K[U_L \rho U_L^\dag,U_L\pi_\rho U_L^\dag]=\K[\rho,\pi_\rho]$}

{\proposition For strategies S2q, S2c, and S3 Condition 1(b) for $\Q$ and $\C$ is satisfied.}

\proof {Under a local unitary we get $\rho\rightarrow U_L\rho U_L^\dag$, $\pi_\rho\rightarrow U_L\pi_\rho U_L^\dag$. To maintain the minimization (maximization) for S2q (S2c) and for S3 the basis of $\pi_\rho$ changes, the measurement in each case, must also changes unitarily: $\M\rightarrow U_L\M U_L^\dag$. Therefore, we have $\Q(U_L\rho U_L^\dag)=\K[U_L\rho U_L^\dag, U_L\M(\rho)U_L^\dag]=\Q(\rho)$ and $\C(U_L\rho U_L^\dag)=\K[U_L\M(\rho) U_L^\dag, U_L\M(\pi_\rho)U_L^\dag]=\C(\rho)$}

{\proposition For strategy S1 Condition 1(b) for $\Q$ and $\C$ may fail.}

\proof{We show this by using a similar example as in the proof for Prop. 3. Let $\rho =\frac{1}{2} \{\ket{00}\bra{00} +\ket{11}\bra{11}\}$ be a state diagonal in the computational basis and fix $\mathcal{M}$ in the computational basis. Now let $U_L$ be a rotation in the $x-z$ plane by an angle $\theta$ for each qubit, i.e. $\ket{0} \rightarrow \cos(\theta)\ket{0} +\sin(\theta)\ket{1}$ and $\ket{1} \rightarrow \sin(\theta)\ket{0} -\cos(\theta)\ket{1}$. Then we get: $\Q(U_L\rho U_L^\dag) =\frac{1}{4} (1+\cos^2(2\theta)) \sin^2(2\theta)$ and $\C(U_L\rho U_L^\dag) =\frac{1}{4}\cos^4(\theta)$. Both $\Q$ and $\C$ are functions of $\theta$}

\subsection{All correlations are positive and total correlations are non-increasing under local operation}
\noindent
The next two conditions turn out to be easy to satisfy, but further restrict the types of function $\K$ can be. Positivity of $\K[\rho_1,\rho_2]\in \mathsf{R}^+$ gives Condition 1(c) for all strategies. 

To ensure Condition 1(d) we can  restrict the function $\K$ to be contractive under a local operation  (without classical communication) $\Lambda_L=\bigotimes_i \Gamma_i$, where $\Gamma_i$ is a generalized operation on the $i$th subsystem.

{\definition Function $\K$ is called a contractive function under a local operation if it satisfies
\begin{equation}\label{contractiveK}
\K[\rho_1,\rho_2] \geq \K[\Gamma(\rho_1),\Gamma(\rho_2)],
\end{equation}
Condition 1(d) is always satisfied when $\K$ is contractive, but this is not a necessary condition.}

\subsection{Classical states have no quantum correlations}
\noindent
{\definition $\M_\chi$ does not change the classically correlated states in its basis $\{\chi\}$: $\M_\chi [ \M_\chi (\rho) ] = \M_\chi(\rho)$.}

Condition 1(e) is always satisfied from  the definition of classical states (Def. 3) and the above (Def. 5).  However, the number of classical states under S1 is much smaller than those of other strategies since they do not remain classical under local unitary transformations as we showed in Prop. 6. 

We have shown above that strategy S1 does not satisfy several of the necessary conditions. All of these conditions are satisfied by rest of the strategies. However, in meeting these criteria we had to restrict $\K$ to be a positive, unitary invariant Eq.~\eqref{unitaryK}, and contractive Eq.~\eqref{contractiveK} under generalized operations. For $\{\M\}$ we use Def. 6 and also assume  the set  $\{\M\}$ contains all the necessary elements required to meet the above conditions. This is true if for example $\{\M\}$ contains (but is not necessarily restricted to) all local orthogonal measurements. We have summarized the results of this section in Table~\ref{table1}.

\begin{table}[t!]
	\begin{center}
\begin{tabular}{cl||c|c|c|c}
N.C.& & $\quad$ S1 $\quad$ & $\quad$ S2q $\quad$ & $\quad$ S2c $\quad$ & $\quad$ S3 $\quad$ \cr
\hline\hline
	 & $\T$ & yes & yes & yes & yes \cr
(a) & $\Q$ & no & yes & yes & yes \cr
	 & $\C$ & no & yes & yes & yes \cr\hline
	 & $\T$ & yes & yes & yes & yes \cr
(b) & $\Q$ & no & yes & yes & yes \cr
	 & $\C$ & no & yes & yes & yes \cr\hline
(c) & & yes & yes & yes & yes \cr\hline
(d) & & yes & yes & yes & yes \cr\hline
(e) & & yes & yes & yes & yes \cr
\end{tabular}
\end{center}
\tcaption{\label{table1}The necessary conditions for various strategies. If we restrict $\K$ to be invariant under local unitary operations Eq.~\eqref{unitaryK} and  contractive Eq.~\eqref{contractiveK} all strategies except S1 meet all the necessary criteria. Additionally we had to put some restriction on the action $\M$ by Def. 6. Above {\emph yes} indicates that the property was proved to hold and {\emph no} indicates that at least one counterexample was presented.}  
\end{table}

\section{Reasonable conditions (continuity of correlations)}\label{RC}
\noindent
Continuity is an important property for any measure on quantum states both from a purely mathematical perspective and a more practical experimental perspective. That is, making small perturbations on a state should not change its correlations by too much~\cite{nielsenEF, donald}. From an experimental perspective, the measurement used to minimize the correlations should be able to get very close to minimizing the correlations in any nearby state. That is, if due to some noise the state is slightly changed, the measurement used to minimize correlations before the perturbation should not give a very different value for the correlations.  A stronger requirement is that the measurement used to minimize correlations before the perturbation should be close to the measurement used after the perturbation. 

We note that the following results  for continuity is independent of how the set   $\{\M\}$ is chosen. In particular the results are valid for both the case when only  orthogonal projective measurements are allowed and for the case where all  POVMs are allowed.

\subsection{Continuity of states and correlations}
\noindent
Suppose we have a state that is slightly perturbed
\begin{equation}\label{pert}
\rho \rightarrow \sigma = (1-\ep)\rho +\ep \tau,
\end{equation}
with $\ep$ being very small. We would like to know how discord changes as the state changes. In fact, as we argued above we would like this change to be continuous.

{\corollary When a state is changed continuously the marginal state also changes continuously:
\begin{equation}
\pi_\rho \rightarrow \pi_\sigma=(1-\epsilon)\pi_\rho+\epsilon\pi_\tau.
\end{equation}}

{\corollary When a state is changed continuously the post-measurement state also change continuously:
\begin{equation} \label{cm}
\M(\rho) \rightarrow \M(\sigma)=(1-\epsilon)\M(\rho)+\epsilon\M(\tau).
\end{equation}}

{\definition We call $\K$  continuous when it satisfies the following inequalities:
\begin{eqnarray}
&\left|\K[\rho_1,\rho_1]-\K[\rho_1,(1-\ep)\rho_2+\ep\tau]\right|\leq f(\ep)\label{eq2},\\
&\left|\K[\rho_1,\rho_2]-\K[(1-\ep)\rho_1+\ep\tau,\rho_2]\right|\leq f(\ep)\label{eq3},
\end{eqnarray}
for all $\rho_2$ and $\rho_2$ with arbitrary $\tau$ and $f(0)=0$. In the following we assume that $\K$ is continuous.}

{\lemma Total correlations are continuous: $|\T(\sigma)-\T(\rho)|\leq g(\ep),$ with $g(0)=0$.}

\proof {We start with $\T(\rho)=\K[\rho,\pi_\rho]$. Let us perturb $\rho$ while keeping $\pi_\rho$ fixed and then the converse. By continuity of $\K$ we have 
\begin{equation}
|\K[\sigma,\pi_\rho]-\K[\rho,\pi_\rho]| \leq f(\ep) \quad {\rm and} \quad |\K[\sigma,\pi_\sigma]-\K[\sigma,\pi_\rho]|\leq f(\ep).
\end{equation} 
Adding the two terms and using the fact that $|A|+|B|\geq|A+B|$ we get 
\begin{equation}
|\K[\sigma,\pi_\sigma]-\K[\rho,\pi_\rho]|\leq g(\ep), 
\end{equation}
where $g(\ep)=2f(\ep)$}

{\theorem{Quantum and classical correlations for strategy S1 are continuous: 
$|\Q(\sigma) - \Q(\rho)| \leq g(\ep)$, with $g(0)=0$.}}

\proof {If state $\rho$ changes according Eq.~\eqref{pert}, using Eqs.~\eqref{cm}-\eqref{eq3} with variation in the first term first and second term second,
\begin{equation}
\left|\K[\sigma,\M(\sigma)] -\K[\rho,\M(\sigma)]\right|\leq f(\ep) \quad {\rm and} \quad
\left|\K[\rho,\M(\sigma)] -\K[\rho,\M(\rho)]\right|\leq f(\ep). 
\end{equation}
Adding the two terms and using triangle inequality, $|A+B|\leq|A|+|B|$, we have 
\begin{equation}
\left|\Q(\sigma)-\Q(\rho)\right|\leq g(\ep),
\end{equation}
with $g(\ep)=2f(\ep)$. The exact argument can be carried out for classical correlations}

{\theorem Quantum (classical) correlations for strategy S2q (S2c) are continuous: $|\Q(\rho) - \Q(\sigma)|\leq g(\ep)$.}

\proof{First note the following inequality: $0\leq \Q_{\M_\sigma}(\rho)-\Q(\rho)$, where subscript $\M_\sigma$ denotes measurement in the basis that minimizes discord for $\sigma$. Without loss of generality we assume $\Q(\rho)>\Q(\sigma)$, which means $0  \leq \Q(\rho)-\Q(\sigma)$. After adding the last two inequalities we have 
\begin{equation}
\Q(\rho)-\Q(\sigma) \leq  \Q_{\M_\sigma}(\rho)-\Q(\sigma) \leq g(\ep).
\end{equation}
We used the result of Thm. 1 to get the final step. If $\Q(\rho)>\Q(\sigma)$ was not true then we simply interchange $\rho$ and $\sigma$ and we get the same result. The same argument can be carried out for classical correlations for strategy S2c}

Thm. 2 only applies to the continuity of quantum (classical) correlations for strategy S2q (S2c). It is, however, unclear if continuity for classical (quantum) correlations still works if we use strategy S2q (S2c). We leave this as an open question. In this respect it is also interesting to ask how $\LL$ changes when we change strategies this is again still open. Further, Thm. 2 does not say anything about the smoothness of correlations. In fact, there are known phenomena that show that quantum discord does not change smoothly~\cite{arXiv:0911.2848, arXiv:1001.5441}. 

{\proposition Continuity  may fail for strategy S3.}

\proof {Once again we rely on an example to prove the proposition. 
Take a state matrix 
\begin{equation}
 \rho= \frac{1}{4} \{\mathsf{I}\otimes\mathsf{I} +\frac{1}{2}\sigma_x\otimes\sigma_z +\frac{1}{2} \sigma_x\otimes\sigma_x\},
 \end{equation} 
which is locally completely mixed  and unitarily transform it into one of three states which are not degenerate locally on the second subsystem using the transformation:
 \begin{equation}
 U_j = \cos\left(\epsilon\right)\mathsf{I}\otimes\mathsf{I} -i\sin \left(\epsilon\right)\sigma_x\otimes\sigma_j
\approx \mathsf{I}\otimes\mathsf{I} -i\epsilon\sigma_x\otimes\sigma_j,
 \end{equation}
with $j=\{x,y,z\}$. The perturbed states
\begin{equation}
\rho_{j}=U_j\rho U_j^\dagger \approx\rho-\frac{i\epsilon}{8} \mathsf{I}\otimes\{[\sigma_j,\sigma_z] +[\sigma_j,\sigma_x]\}
\end{equation}
are $\epsilon$ close to each other.
\begin{equation}\label{p7s}
\rho_{x} =\rho-\frac{\epsilon}{4} \{\mathsf{I}\otimes\sigma_y\}, \;
\rho_{y} =\rho+\frac{\epsilon}{4} \{\mathsf{I}\otimes(\sigma_z-\sigma_x)\}, \;
\rho_{z} =\rho+\frac{\epsilon}{4} \{\mathsf{I}\otimes\sigma_y\}. 
\end{equation}

For most reasonable types of S3 measures the discord will be very different for the above states because $\rho$ is more sensitive to decoherence in the $y$ direction. For bipartite $2\times 2$ systems, an asymmetric measure like the \emph{measurement induced non locality} (MINL) has a simple analytical expression~\cite{MINL}. In this case $\K[\rho,\M(\rho)]=|\tr(\rho^2)-\tr[\M(\rho)^2]|$ and $\M$ is an orthogonal rank-1 projective measurement on system A.  The resulting MINL is given by: $N(\rho)=\frac{1}{32}$, $N(\rho_{x,z})=\frac{1}{32}$, and $N(\rho_{y})=\frac{1}{64}$. This is independent of $\epsilon$ for arbitrarily small $\epsilon$}

Another example for discontinuity in S3 using MID and Werner states can be found in~\cite{arXiv:0905.2123}.

\subsection{Strong continuity}
\noindent
We may wonder if the measurement basis, and consequently the classical state, changes continuously under a perturbation: We call this \emph{strong continuity of $\M$} (SCM). This will not be an issue with strategy S1, since SCM is enforced by keeping $\M$ fixed. Strong continuity would be nice but may not be reasonable condition (rather desirable) as we are interested in the correlations not the specifics of $\M$. 

{\proposition Strong continuity of $\M$ (and consequently the continuity of the classical state) does not hold in general for correlation measures for strategies S2q, S2c, and S3. SCM fails for all well-known measures.}

\proof{ We construct a proof by example again. Take a state that has marginals very close to the maximally mixed state
\begin{equation} 
\rho=\ket{\psi^+}\bra{\psi^+},
\end{equation}
where $\ket{\psi^+}=\frac{1}{\sqrt{2}}\left[\ket{00}+\ket{00}\right]$ and perturb it in the direction of the state $\ket{\phi\phi}$ with $\phi$ arbitrary
\begin{equation}
\sigma_\phi= (1-\ep)\ket{\psi^+}\bra{\psi^+}+ \ep\ket{\phi\phi}\bra{\phi\phi}.
\end{equation}

After a measurement in some basis $\hat{m}$, on the first system only, we will get 
\begin{eqnarray}
\M_{\hat{m}}(\rho_{\hat{n}}) &=&
\frac{1-\ep}{2} \left[ \ket{00}_{\hat{m}}\bra{00}_{\hat{m}} +\ket{11}_{\hat{m}}\bra{11}_{\hat{m}} \right]\nonumber\\ 
&&+\ep \left( p_0 \ket{0}_{\hat{m}} \bra{0}_{\hat{m}} \otimes\ket{\phi}\bra{\phi}
+ p_1 \ket{1}_{\hat{m}} \bra{1}_{\hat{m}} \otimes\ket{\phi}\bra{\phi}\right),
\end{eqnarray} 
where $p_0$ and $p_1$ are the probabilities of the respective measurement outcomes.  It is easy to see that  for the more standard choices of $\K$ and $\M$ the discord is minimized for measurement  in the basis of $\ket{\phi}$. If we take two states $\sigma_{\hat{x}}$ and $\sigma_{\hat{y}}$, both $\ep$ close to $\rho$ and therefore each other, we will get the minimum discord in two very different basis, and therefore two very different classical states in basis $\hat{x}$ and $\hat{y}$ respectively. However, these two states, $\M(\sigma_{\hat{x}})$ and $\M(\sigma_{\hat{y}})$, are not close to each other under certain choices of $\K$:  mutual information, conditional entropy, relative entropy, Hilbert-Schmidt distance. In fact, for this example discord, relative entropy of discord, and MID are all equivalent to each other. This means that SCM fails for all well-know measures: discord, relative entropy of discord, MID, geometric discord, MINL}

A side note: the states above ($\rho$ and $\sigma$) are not connected to each other by a unitary transformation. However, a similar argument can be made for unitary perturbations,  the calculations is more messy but gives the same results. In fact, for S3 the example of Prop. 7, where the perturbation is unitary, shows that the local eigenbasis of the second subsystem in $\rho_y$ is very different from the that of the other two giving a discontinuity in SCM.  

\subsection{Weak continuity}
\noindent
If strong continuity is violated, we would still like the correlations of the perturbed state to be close to the correlations of the unperturbed state, even if the wrong basis of measurement is applied to measure the correlations: We call this \emph{weak continuity of $\M$} (WCM). Suppose two quantum states are very close to each other but the measurement basis that minimize respective discords are very different from each other. Then we ask, does discord change drastically if we minimize using the wrong basis?

{\theorem Weak continuity of measurement basis is satisfied for S2q (S2c) $\Q_{\M_\sigma}(\rho)-\Q(\rho)\leq h(\ep)$ ($\C_{\M_\sigma}(\rho)-\C(\rho)\leq h(\ep).$).}

\proof{Adding the final results of Thms. 1 and 2 and using the triangle inequality, $|A+B|\leq|A|+|B|$, we get
\begin{eqnarray}
\left|\Q(\rho)-\Q(\sigma)\right|+\left|\Q(\sigma)-\Q_{\M_\sigma}(\rho)\right| &\leq& 2 g(\ep)\nonumber\\
\left|\Q(\rho)-\Q(\sigma)+\Q(\sigma)-\Q_{\M_\sigma}(\rho)\right| &\leq& 2 g(\ep)\nonumber\\
\left|\Q(\rho)-\Q_{\M_\sigma}(\rho)\right| &\leq& h(\ep),
\end{eqnarray}
where $h(\ep)=2g(\ep)$.
The same logic will give the same results for $\C$}

{\proposition WCM  may fail for strategy S3.}

\proof{ Using the example of Prop. 7 Eqs. \eqref{p7s} one can see that WCM fails}

\begin{table}[t!]
\begin{center}
\begin{tabular}{cl||c|c|c|c}
R.C. & & $\quad$ S1 $\quad$ & $\quad$ S2q $\quad$ & $\quad$ S2c $\quad$ & $\quad$ S3 $\quad$ \cr
\hline\hline
	&$\T$ & yes & yes & yes & yes \cr
(a) &$\Q$ & yes & yes & ?   & no \cr
	&$\C$ & yes & ?   & yes & no \cr\hline
(b) &     & enforced  & no  & no & no\cr\hline
(c) &     & -   & yes & yes & no \cr
\end{tabular}
\end{center}
\tcaption{\label{table2}Continuity using various strategies: S2q and S2c are continuous and WCM,  S3 is neither. All strategies except the trivially continuous S1 fail strong continuity. Above {\emph yes} indicates that the property was proved to hold, {\emph no} indicates that at least one counterexample was presented, and \emph{?} indicates that neither was found.}
\end{table}
We have summarized the results of this section in Table~\ref{table2}. We should also point out that the analytical results in this and previous section apply just as well for the continuous variable case~\cite{arXiv:1003.3207, arXiv:1003.4979}. 

\section{Debatable conditions}\label{DC}

\subsection{Conditions for bipartite pure states}
\noindent
For a pure bipartite state we can use the Schmidt decomposition $\ket{\Psi}=\sum_i\sqrt{\lambda_i} \ket{\psi_i^A\phi_i^B}$ where $\{\ket{\psi_i^A}\}$ and $\{\ket{\phi_i^B}\}$ form an orthonormal basis on their respective spaces. The marginal states $\rho^A=\sum_i\lambda_i \ket{\psi_i^A}\bra{\psi_i^A}$ and $\rho^B=\sum_i\lambda_i \ket{\phi_i^B}\bra{\phi_i^B}$ are defined by the Schmidt coefficients $\lambda_i$. It is then natural to expect that the correlations will also be defined by these coefficients. In which case they can be determined by some function on the local reduced marginals. Here we note that this makes it very natural to assume $\M_\Psi$ does not change the marginals for pure states. Indeed for most types of discord on pure states S2 and S3 give the same measurement basis. In fact, for bipartite pure states the entropic discord is same as the entanglement measure, i.e. entropy of one of the reduced state. This does not hold true for multipartite systems~\cite{arXiv:0911.5417}.

\subsection{Additivity of correlations}
\noindent
It is reasonable to expect the total correlations are the sum of classical and quantum, correlations.  We can do this  either by  imposing constraints on $\K$ and $\M$, or by redefining one type of correlation as a function of the others. A byproduct of this property is that S2q and S2c are equivalent.

{\theorem If $\T=\C+\Q$ then S2q and S2c are equivalent.}

\proof {Since $\T$ is the same for all strategies minimizing $\Q$ will maximize $\C$ and vice versa}

One can also think of some correlations being simultaneously quantum and classical leading to an inequality 
\begin{equation}
\T\le\C+\Q\le2\T.
\end{equation}
We call the last condition 3(b$'$). Where we assume that neither the classical nor the quantum correlations can exceed the total correlations (see Sec.~7 for an example). Another,   weaker version of this condition has an equality for bipartite pure states only.  We will call this condition 3(b$''$): $\T=\C+\Q$ for pure states. These conditions are only of interest if all three quantities have the same  dimensions. 

\subsection{$\C$ and/or $\Q$ are non increasing under local operations}
\noindent
In~\cite{hv} the above condition was presented as a necessary condition for $\C$. With the addition of classical communication it is also a condition for entanglement measures. Since $\C$ can (classically) be increased under communications, it is clear that the condition requiring that $\C$ is non-increasing under LOCC is unreasonable. $\Q$ can both increase and decrease under local operations in most measures of correlations. For this reason there is no analogue of the distillation process which is used in entanglement theory. Two weaker  versions of the property  read:

3(c$'$) - {\it $\C$ and/or $\Q$ are non increasing under partial trace.} 

3(c$''$) - {\it $\C$ and/or $\Q$ are non increasing under the addition of a subsystem.} 

The importance of 3(c$''$) is shown by a remarkable application given in~\cite{arXiv:1112.3967}. There it is shown that if a measure of correlation is monogamous and satisfies conditions 1(b), 1(c), and 3(c$''$) then this measure vanishes for all separable states.

{\theorem If $\{\M\}$ includes all local POVMs then $\C$ is non-increasing under local operations $\Lambda_L$ for S2c.}

\proof{Any local operation followed by a measurement could be replaced by a different measurement $\M(\Lambda_L (\rho) ) = \M_{\Lambda_L} (\rho)$. Since we are optimizing over all possible measurements a local operation could not increase the classical correlations}

One might also think that we can increase classical correlations at the expense of quantum correlations using local operations. It is not an unreasonable assumption but it is tricky since it suggests possible violations of causality. Using S3 it is possible to show that we can have an interpretation of classicality increasing at the expense of quantumness.

{\proposition For S3 local operations can increase classical correlations at the expense of quantum correlations. }

\proof{We prove this with an example: Starting with the state 
\begin{equation}
\rho=\frac{1}{4}\{\mathsf{I}\otimes\mathsf{I}+\ep(\mathsf{I}\otimes\sigma_z+\sigma_z\otimes\mathsf{I})+c\sigma_x\otimes\sigma_x\},
\end{equation}
with very small $\ep$. The measurement basis for S3 will be in the $\hat{z}$ basis on both sides and the resulting state will be
\begin{equation}\label{afterMID}
\M(\rho)=\frac{1}{4}\{\mathsf{I}\otimes\mathsf{I}+\ep(\mathsf{I}\otimes\sigma_z+\sigma_z\otimes\mathsf{I})\}
\end{equation} 
Which is an uncorrelated state so $\C(\rho)=0$. 

However, if both parties were to perform a local measurement in the $\hat{x}$ basis and with some very low probability $\ep$ locally turn their state to a $+$ eigenstate of $\sigma_x$ they will have a state.
\begin{equation}
\rho_{lo}=\frac{1}{4}\{\mathsf{I}\otimes\mathsf{I}+\ep(\mathsf{I}\otimes\sigma_x+\sigma_x\otimes\mathsf{I})+c\sigma_x\otimes\sigma_x\} 
\end{equation}
In which case we get the same total correlations (up to $\ep$) but the S3 discord will be zero since the measurement basis would be $\hat{x}$  so the correlations are all classical.   Using local operations and no communications we gain an increase in classical correlations at the expense of quantum correlations}
 
\subsection{Symmetry under interchange of the subsystems}
\noindent
The set of measurements $\{\M\}$ can sometimes be asymmetric with respect to the subsystems, for example it could be a set of local measurements on only one subsystem. In this case swapping the subsystems around may change the classical and quantum correlations. It is often natural to assume that the correlations do not depend on which subsystem is being observed. However, if the correlations are related to the ability to perform some task, this assumption can be inappropriate. In this respect some choose to classify bipartite systems as being classical-classical, quantum-classical, classical-quantum, and quantum-quantum. See~\cite{arXiv:1105.2768} for such an application of classical-quantum correlated state.

\subsection{Relation to entanglement measures}
\noindent
The criteria proposed in this section are  based on the criteria for entanglement measures. It is important to stress that the general method described here cannot be used to describe entanglement measures. In general the operation $\M$ would have to be non completely positive if we wanted to use it for a measure of entanglement~\cite{arXiv:1105.4115, arXiv:0707.2195}. 

Another requirement may be added so that quantum correlations should be greater than  or equal to entanglement measured using the same type of metric. For instance, the relative entropy of discord is always greater than or equal to relative entropy of entanglement~\cite{arXiv:0911.5417}. If we define entanglement in the same way as we defined generalized discord, i.e. the distance to the closest separable state then this will always be satisfied, as the closest classical state is also a separable state.
It is not reasonable to expect this kind requirement to hold  when comparing two different types of correlations, e.g. in~\cite{PhysRevA.77.042303} an example is given where entanglement of formation is greater than quantum discord. On the other hand we can also require that when $\Q>0$, $\C>0$ (quantum correlations cannot exist without classical correlations). This is always the case with quantum discord, however relative entropy of discord (see Sec.~7) and quantum zero-way deficit do not satisfy this. An example where quantum correlations exist even when classical correlations are absent is given in~\cite{arXiv:0705.1370}.

Entanglement can be defined operationally in terms of formation, distillation, etc. These notions do not exists for generalized quantum discord. On the other hand quantum discord and other similar measures are intimately related to entanglement. Some of these relationships arise when considering the mixed state as a part of some larger pure state. These relationships between discord and entanglement translate some of the properties of entanglement to  discord. Below we give a brief review of such relationships.

One of the strongest relations between entanglement and discord is given by the Koashi-Winter formula~\cite{KoashiWinter}. Using this formula a conservation law for discord is derived in~\cite{PhysRevA.84.012313} and discord is shown to be equivalent to the consumption of entanglement in a protocol called extended state merging~\cite{PhysRevA.83.032324}. Further regularized version of discord is shown to be the difference in entanglement cost and entanglement of distillation~\cite{arXiv:1007.0228}. In two independent but similar studies quantum discord, quantum deficit, and relative entropy of discord are related to entanglement generation in measurement and entanglement activation~\cite{PhysRevLett.106.160401, PhysRevLett.106.220403}. Based on monogamy of entanglement, monogamy of discord is analyzed in~\cite{arXiv:1108.5168, arXiv:1109.1696v1}. From an operational point of view, manipulating quantum correlations may require entanglement in some scenarios~\cite{BTgates}

\subsection{Relation to criteria for genuine multipartite correlations}
\noindent
The general method presented above can be used for quantifying multipartite correlations, however it may not be ideal for capturing {\it genuine} multipartite correlations. That is, given some value for the classical or quantum correlations, there is very little information about the  number of states that play an active part in these correlations. Bennett et al.~\cite{PhysRevA.83.012312}  present three criteria for genuine multipartite correlations and three  criteria for quantifying the degree of correlations in terms of the number of parties that play (a genuine) part in the correlations. Of these criteria one is similar to the criteria presented above. {\it Local operations and postselection cannot increase the degree of correlations} which can be compared with criteria 3(c).  However this criteria is a stricter vesion of 3(c) which reads {\it Local operations and postselection cannot create correlations} in other words correlations cannot be increased if they are zero unless some communication is used.  Even this stricter version does not hold for all measures. It usually fails for quantum correlations since they can be created locally from classical correlations.  However it always holds for total correlations due to 1(a) and 1(c).

\section{Compatibility of various methods and correlation measures}
\noindent
Without having gone into details about the exact form of $\K[\rho,\M(\rho)]$ we showed that S1 and S3 fail to meet some of the requirements to be measures of quantum and classical correlations. S1 fails to meet even the minimum requirements of local unitary invariance, and zero discord for product states. S3 is at some points not continuous and  does not follow the weak continuity for measurement. Interestingly S2 is only weakly continuous for $\M$. The requirement that classical correlations do not increase under local operations does not hold either for S3.  Below we briefly discuss some of the well known types of discord in light of the results above.

\subsection{Forms of $\K$ and $\M$}
\noindent
There are (infinitely) many different functions $\K$ one can choose to define a measure for correlations. In~\cite{arXiv:1105.4920}, the authors describe different types of entropic-discords by using different functions for $\K$.
\begin{itemize}
\item[(a)] Mutual information $K_I[\rho,\M(\rho)]=|I(\rho)-I(\M(\rho))|$; 
\item[(b)] Conditional entropy $K_D[\rho,\M(\rho)] =|S(\rho^B|A)-S(\M(\rho)^B|A)|$; 
\item[(c)] von Neumann entropy $K_S[\rho,\M(\rho)]=|S(\M(\rho))-S(\rho)|$;
\item[(d)] Geometric $K_G[\rho,\chi]=\|\rho-\M(\rho)\|_2.$
\end{itemize}
We have added another option to the list above, option (d). This is a geometric measure based on the Hilbert-Schmidt distance: where $\|\chi\|_2=\tr[\chi\chi^\dagger]$~\cite{arXiv:1004.0190}. The absolute values in (a-c) are unnecessary since the arguments are always positive, we keep them as a formality  to comply with Def. 1. The functions (a) and (b) are defined for bipartite states while (c) and (d) can be used for  multipartite correlations. 

We adopt the  requirement $\M_\chi(\M_\chi(\rho))=\M_\chi(\rho)$  of  (Def. 6) which limits our choice of $\M$ to orthogonal projective measurements\footnote{One might come up with some set of POVMs which will meet the requirement $\M_\chi(\M_\chi(\rho))=\M_\chi(\rho)$ for all $\rho$ and $\chi$ but it is unlikely that such a set will meet the necessary conditions or produce meaningful classical states.}. However in the entropic cases (a-c) , a POVM can be achieved using  Neumark's extension and the results remain equally valid. In general measurements can be of different ranks, but usually rank-1 measurements will give the desired results (see~\cite{arXiv:1105.4920} for a detailed discussion). There is however more freedom in what kind of measurements are allowed, for example measurements on only some (or one) of the subsystems, or measurements on one system first and then on another etc.  In any minimization procedure S2q and S2c we must assume that the minimization is done by an ``all knowing" observer. If this is not the case we must restrict $\M$ or $\K$ or use a different strategy such as S1 or S3. 
 
Restricting to orthogonal measurements  has fundamental consequences.  In the case where a minimum can  be achieved by a non-orthogonal rank-1 POVM, one may introduce an ancilla system and work with orthogonal measurements. In this case the correlations may be changed simply by adding the new ancilla system. However this is acceptable  if we drop condition 3(c), since it can only increase classical correlations at the expense of quantum correlations, the total correlations $\T$ will remain unaffected.

\subsection{A short discussion on known measures}\label{measures}
\noindent
We can now look at some measures used in recent literature. The measurement set $\{\M\}$ here is a set of POVMs, usually but not necessarily  rank-1 orthogonal measurements.  First we point out which function the measure is based on and which strategic category it belongs to. When necessary we will discuss the form of $\{\M\}$. 

\subsubsection{Discord (D)} 
\noindent
The  quantum discord~\cite{arXiv:quant-ph/0011039, hv, arXiv:quant-ph/0105072} is defined using $\K_D$ and strategy S2q or S2c. For orthogonal measurements this is the same as using $\K_I$  and both S2q and S2c give the same results. In general one may add a pure ancilla as the initial state to get the same result for general POVMs.

Discord meets all the necessary conditions as shown in the early papers on the subject.   It also meets the  reasonable (continuity) conditions, except 2(b) (SCM). To prove continuity  we can use Fannes' inequality~\cite{NC} for continuity of entropy
\begin{equation}
S(\rho)-S(\sigma)\leq \|\rho-\sigma\|_1\log d+h(|\rho-\sigma\|_1).
\end{equation}
in our general proofs  for  continuity, here $\|\rho\|_1=\tr[\sqrt{\rho\rho^\dagger}]$ is the trace distance. 

Additionally, conditional entropy  is defined by the Schmidt coefficients for pure states so 3(a) is satisfied.  Condition 3(b) is satisfied since the conditional entropy of $\pi$ is the entropy of the reduced marginal that has not been measured, from Thm. 4  S2q and S2c are equivalent for discord.    Cond. 3(c) is satisfied for classical correlations~\cite{hv}, but not for quantum correlations~\cite{BTgates}. Finally, 3(d) is not satisfied since this is an asymmetric measure  of bipartite states (meaning $\{\M\}$ acts on one party only). 

Discord may also be defined using strategy S1;~\cite{arXiv:quant-ph/0105072}. Due to the nature  of mutual information, this measure still satisfies Cond. 1(a), but fails for 1(b).  Since it is S1, it satisfies the reasonable conditions for continuity. Finally, it is the same as the S2 discord for  the conditions 3(a-d).

\subsubsection{Measurement induced disturbance (MID)}
\noindent
Using $\K_I$, strategy S3 and measurements on all subsystems Rajagopal and Rendell and independently Luo~\cite{PhysRevA.66.022104, PhysRevA.77.022301} introduced MID. It satisfies all necessary conditions. But since it is a S3 measure, it fails all continuity conditions, as we have shown using several examples above. For bipartite pure states, MID is the same as the  discord. Therefore, it satisfied 3(a). It satisfies all other debatable conditions, except 3(c) since both quantum and classical correlations can be increased using local operations. It is also a symmetric measure satisfying Cond. 3(d).  An asymmetric version of MID (equivalent to the S3 discord) was introduced in~\cite{BTdemons}.

\subsubsection{Relative entropy of discord (RED)} 
\noindent
RED~\cite{arXiv:quant-ph/0112074, arXiv:quant-ph/0202123, arXiv:0911.5417} is defined using $\K_S$ and S2q. This can be a symmetric or asymmetric measure. RED satisfies all necessary and reasonable conditions, except 2(b). For bipartite pure states, RED is the same as the  discord. Therefore, it satisfies 3(a). For 3(b), we have $\T+\LL=\Q+\C$. 
It satisfies 3(d) when the symmetric version is considered, but otherwise not. 

Another equivalent way of defining RED is using the (lowest) relative entropy from the state $\rho$ to a classical state. Surprisingly the optimal  classical state is given by an orthogonal projective measurement on $\rho$~\cite{arXiv:0911.5417}.

\subsubsection{Geometric discord (GD)}
\noindent
GD is defined using $\K_G$ and S1, S2q or S2c with either a symmetric or asymmetric measurement. When S1 is considered, conditions 1(a) and (b) may fail. They will not fail for S2q or S2c. For S1, S2q, and S2c, all other necessary conditions and the conditions for continuity are satisfied by GD.  It also satisfies the additivity condition for bipartite pure states 3(b) and therefore 3(a) as well (see below). Using Thm. 5  condition 3(c) is satisfied for classical correlations if we use S2c. The geometric measures have the advantage of an analytic formula for bipartite states~\cite{arXiv:1004.0190, arXiv:1010.1920}. 

{\proposition The geometric discord  of bipartite pure states satisfies conditions 3(a) and 3(b).}

\proof{Let $\rho=\sum_{ij} \sqrt{\lambda_i\lambda_j} \ket{\psi^A_i\phi^B_i}\bra{\psi^A_j\phi^B_j}$. Then the corresponding classical state is $\M(\rho)=\sum \lambda_i \ket{\psi^A_i\phi^B_i} \bra{\psi^A_i\phi^B_i}$, $\pi\rho=\pi^A\otimes\pi^B=\sum \lambda_i \lambda_j \ket{\psi^A_i} \bra{\psi^A_i}\otimes \ket{\phi^B_j} \bra{\phi^B_j}$. By direct computation of different correlations we have
\begin{eqnarray}
\T&=& \tr \left[ (\rho - \pi_\rho)^2 \right] =1 + \tr(\pi^2_\rho)-2\sum_i\lambda_i^3\\
\Q&=& \tr \left[ (\rho - \M(\rho))^2 \right] =1 - \tr \left[ \M(\rho)^2 \right]\\
\C&=& \tr \left[ (\M(\rho)-\pi)^2 \right]
=\tr \left[ \M(\rho)^2 \right] + \tr \left(\pi_\rho^2 \right) - 2\sum_i\lambda_i^3.
\end{eqnarray}
Each function above is a function of $\lambda_i$, therefore 3(a) is satisfied. Adding the last two functions gives the first: $\T=\Q+\C$}

In all references to the geometric discord we only consider S2q and S2c. We note that the classical state obtained by $\M(\rho)$ is also the closest classical state overall~\cite{shunlonggeo}, i.e using  $\min_{\chi}|\rho-\chi|$ where $\chi$ is a classical state. 

\subsubsection{Measurement induced nonlocality (MINL)} 
\noindent
MINL ~\cite{MINL} is a generic asymmetric S3 strategy with an optimization process for degenerate marginals.  However this optimization process does not overcome the discontinuities associated with S3.  One may use  $\K_G$  to obtain an analytic formula. In general it meets the same criteria as MID except symmetry. 

The results of this section are summarized in Table~\ref{table3}.
 
\begin{table}[t!]
\center{\begin{tabular}{ll||c|c|c|c}
 & & $\quad$ D $\quad$ & $\quad$ MID/MINL$\quad$ & $\quad$ RED $\quad$ & $\quad$ GD $\quad$ \cr
\hline\hline
N.C. &     & yes  & yes & yes  & yes    \cr\hline
R.C. & (a) & yes  & no  & yes  & yes    \cr
	 & (b) & no   & no  & no   & no     \cr  
	 & (c) & yes  & no  & yes  & yes    \cr\hline
D.C. & (a) & yes  & yes & yes  & yes    \cr
     & (b) & yes  & yes & yes  & yes  \cr
     & (c) & $\C$ & no  & $?$ & $\C$   \cr
     & (d) & no   & yes/no & yes/no & yes/no \cr
\end{tabular}}
\tcaption{\label{table3} Comparing some known versions of discord using the criteria for correlations. Above \emph{yes} means it is satisfied and \emph{no} means we gave a counterexample in the previous sections. For 3(b) either 3(b) or 3(b$'$) is satisfied. For 3(c) $\C$ means it is satisfied only for classical correlations. For 3(d) \emph{yes/no} means \emph{yes} for the symmetric version and \emph{no} for the asymmetric version.}  
\end{table}

\section{Demons, basis, and continuity}
\noindent
Maxwell's demon and work deficit paradigms are some of the most interesting physical interpretations of quantum correlations  beyond entanglement. These paradigms often involve some limitations on what local demons can do, usually by restricting their communications~\cite{arXiv:quant-ph/0410090}. However we can also restrict the knowledge of the demons~\cite{BTdemons}  and/or their apparatus. In the following we do both, first we show that when the demons lack knowledge the best strategy might be S3, and that there is some physical basis for the non-continuous nature of these measures. Next we examine the situation when there is imperfect knowledge of the state due to some uncertainties and/or the apparatus is imperfect. In this case, which is close to the realistic situation in a lab, we see that continuity is necessary to make any sense of correlations. 

\subsection{Basis agreement}\label{demonexample}
\noindent
Imagine the following scenario for Szilard's engine~\cite{RevModPhys.81.1, Szilard}. Alice and Bob share a quantum state $\rho$ but each only knows his own density matrix. They must extract as much work out of the system as possible with no communication (they can communicate later to maximize the efficiency of erasing their records).  The difference between the amount of work Alice and Bob can generate and the amount of work a nonlocal all knowing  demon can generate is related to the MID~\cite{BTdemons, arXiv:1105.4920}, but this only applies in the case where the local states are not degenerate. To understand this we note that the best measurement strategy in this case is a measurement in the local eigenbasis. Since Alice and Bob have no knowledge the total state their best guess would be to maximize their own work. If there are degeneracies  however, there is some ambiguity as to which is the best measurement basis, and they must choose some arbitrary direction. In the worst case scenario the work is related to the MINL, because in MINL a maximization is required when the local basis are unknown. 

\subsubsection{Example 1: Ambiguous local basis}
\noindent
Let the system be in a classical state $\rho=\frac{1}{2}\{\ket{00}\bra{00} +\ket{11}\bra{11}\}$. The reduced states are maximally mixed, and therefore there is no well-defined local basis. In the best case scenario the measurement is made in the computational basis and the maximal amount of work is extracted from the engine giving a work deficit of 0. The reason they can get any work out of the system, once all accounting has been done, is that they can use the correlations to improve the efficiency of the erasure process. In the worst case scenario there are no correlations and the total amount of work gained is zero giving a work deficit of $k_bT$. However, in the slightly modified case of $\rho=\frac{1}{2}\{ (1+\ep) \ket{00}\bra{00} +(1-\ep) \ket{11}\bra{11}\}$, there is no ambiguity and the work deficit is 0. This is an example of a possible interpretation for MID despite the discontinuity. 

\subsubsection{Example 2: No communication with prior knowledge}
\noindent
In this respect it is interesting to note that if Alice and Bob are not allowed to communicate but have some prior knowledge of the state, they can choose the optimal strategy in all cases, removing the ambiguity~\cite{BTdemons}. In fact the best strategy is often not to measure in the eigenbasis but in some other basis. Moreover if Alice and Bob were allowed to discuss their strategy ahead of time they could get more work out of the system. For example, if they shared a singlet state it would be better if they both measured in the same basis, although that basis is completely arbitrary. In the case when Alice measures in the $z$ basis and Bob in the $x$ basis, they wind up with maximally mixed state rather than a  maximally classically correlated  state.

\subsection{Choosing a basis in symmetric and  asymmetric measures}
\noindent
The asymmetric versions of discord have been justified in a number of scenarios like state merging~\cite{PhysRevA.83.032324, arXiv:1008.4135} and work deficit~\cite{arXiv:quant-ph/0410090, arXiv:quant-ph/0202123}. It is interesting to note that when optimizing, the optimal basis used by one side and the optimal basis used by the other side can be different than the optimal basis used by both even when no communication is allowed. 

\subsubsection{Example 3: Symmetric measures and prior communication}
\noindent
We give the following example using Hilbert-Schmidt distance as $\K$ and rank-1 orthogonal projective measurements as $\M$. Again we define  $\ket{\psi^+}=\frac{1}{\sqrt{2}}\left[\ket{00}+\ket{11}\right]$.
\begin{equation}
\rho=(1-c)\ket{\psi^+}\bra{\psi^+}+c\ket{0+}\bra{0+},
\end{equation}
in which case the optimal strategy for an asymmetric measure would be measure in the $\hat{z}$ ($\hat{x}$) basis for a measurement on A (B). But a symmetric measure would be optimized if both used the same basis. 

\subsection{Imperfect demons}
\noindent
One of the main reasons to require  correlations to be continuous and to have upper bounds on how they change for nearby states is that in real life  experimental apparatus is never perfect. For example when tomography is used to determine the state, any small uncertainties could cause a big difference in the amount of work that can be extracted from the system. This is clear in the examples given in Sec.~8 and in (Prop. 7). The S3 measure of work deficit in this case would not be a good measure of the quantumness of the state. 

Another example is where the demon requires to run some kind of numerical approximation in order to find the optimal strategy. Here again continuity will be of major importance as will the weak continuity of the measurement basis. 

\section{Conclusions}
\noindent
The nature of quantum correlations makes them hard to quantify in a consistent manner. Different physical scenarios lead to different measures of correlations. It is however interesting to ask if a measure which is associated with a certain physical paradigm is indeed a measure of correlations.  We built a set of criteria to judge all possible measures of correlations. Of these criteria five were identified as more fundamental than the rest. A physical quantity which fails to meet them cannot be considered  a `proper' measure of correlations. The various measures of discord used in recent literature were found to be consistent with these criteria.  

Continuity on the other hand is more elusive, no measure of correlations was found to be compatible with SCM and the S3 based measures fail all continuity criteria. The measures of discord with S2 such as discord, RED, and geometric discord are however continuous and WCM. As are the set of S2 based measures based on a continuous discord function $\K$. This is the first proof of continuity for measures of quantum correlations beyond entanglement.  

The last set of criteria dubbed {\it debatable} are related to the interplay between classical and quantum correlations. Surprisingly most measures of correlations meet most of these criteria or at least some variation on them. We leave the question of their relevance open. 

Of the known measures of correlations, discord, relative entropy of discord, and other entropic discord are compatible with the majority of requirements. A symmetric version of discord using orthogonal measurements and $\K_I$ is consistent with all requirement and the symmetric versions of the other two is consistent with a slightly weaker version of 3(b) and with all the others (with the possible exception of 3(c) for S2c measures). We did not find a measure which satisfies SCM,  such a measure might exist and make sense in some scenarios. One may try to use a different optimization strategy than those presented above to overcome this obstacle. From a physical point of view, SCM makes the classical state $\M_\rho(\rho)$ associated with $\rho$ more relevant. However, this has little bearing on the quantifiers of the correlations. They remain relevant even when SCM fails. 

While the search for measures of quantum, classical, and total correlations continues it is clear that some measures should not be considered as quantifiers of correlations. The criteria presented in this paper should be used as a general guideline for measures of correlations. The points at which some of these criteria fail for some measures might also have some physical relevance as in the highly symmetric states where continuity  and WCM for S3 breaks down. 

\nonumsection{Acknowledgements} 
\noindent
We would like to thank S. Luo, T. Paterek, D. Terno, and V. Vedral for their helpful comments and suggestions. AB would like to thank CQT for their hospitality. KM acknowledges the financial support by the National Research Foundation and the Ministry of Education of Singapore and is grateful to the Dept. Physics at the Macquarie University for their hospitality.

\nonumsection{References}
\noindent
\bibliographystyle{is-unsrt}
\bibliography{rmp1}
\end{document}